\documentclass[aps,prl,twocolumn]{revtex4}
 \usepackage[usenames,dvipsnames]{color}
\usepackage{graphicx}
\usepackage{dcolumn}
\usepackage{bm}
\usepackage{amsmath}
\usepackage{gensymb}
\usepackage{amsfonts}
\usepackage{ragged2e}
\usepackage{psfrag}
\usepackage[pdftex,bookmarks,colorlinks=false]{hyperref}

\begin{document}
\title{High-Q/V two-dimensional photonic crystal cavities in 3C-SiC}

\author{Ioannis Chatzopoulos}
\author{Francesco Martini}
\altaffiliation{Now at: Istituto di Fotonica e Nanotecnologie (IFN), CNR, via Cineto Romano 42, 00156 Roma, Italy}
\author{Robert Cernansky}
\author{Alberto Politi}
\email{A.Politi@soton.ac.uk}
\affiliation{School of Physics and Astronomy, University of Southampton, Southampton, SO17 1BJ, United Kingdom}

\begin{abstract}
Solid state quantum emitters are between the most promising candidates for single photon generation in quantum technologies. However, they suffer from decoherence effects which limit their efficiency and indistinguishability. For instance, the radiation emitted in the zero phonon line (ZPL) of most color centers is on the order of a few percent (e.g. $NV^-$ centers in Diamond, $V_{Si}V_{C}$ in $SiC$) limiting the emission rate of single photons as well as the efficiency. At the same time, reliable interfacing with photons in an integrated manner still remains a challenge on both diamond and SiC technology. Here we develop photonic crystal cavities with Q factors in the order of 7,100 in 3C SiC. 
We discuss how this high confinement cavity can significantly enhance the fraction of photons emitted in the ZPL and improve their characteristics. In particular, the increased efficiency and improved indistinguishability can open the way to quantum technologies in the solid state.
\end{abstract}

\maketitle

\section{Introduction}
\label{sec:1}

The field of quantum science and technology offers a unique opportunity for studying the fundamentals of physics while delivering technological innovation. Still, many challenges remain unresolved. The ideal platform for implementing quantum technology experiments would combine a list of features\cite{ladd2010quantum,divincenzo2000physical} that has been proven hard to satisfy. Among them, the most challenging is the requirement of scalability. Considering this limitation, solid state quantum systems coupled to integrated optical circuits have attracted an increasing amount of interest. To benefit from their capabilities in quantum optics experiments, controllable single photon sources, possibly coupled to a long lived spin state, are required to display high level of indistinguishability while maintaining high efficiencies. Despite recent advancements\cite{lodahl2015interfacing,somaschi2016near}, an all inclusive technology remains to be found. 

Platforms in solid state have attracted interest, thanks to their more robust spin control and high brightness. During the last decade, one of the most studied platforms in solid state have been color centers in diamond. In particular, the nitrogen vacancy ($NV^-$) center exhibits electronic spins with millisecond coherence times\cite{aharonovich2016solid,gao2015coherent,balasubramanian2009ultralong} that can be coherently manipulated with optical and microwave pulses. While remarkable attempts have been made to interface $NV^-$ centers with integrated circuitry \cite{wan2018two,burek2014high,faraon2012coupling}, the fabrication of diamond photonic components remains challenging. Also, the considerable cost of the material along with the optical wavelength of the $NV^-$ center ZPL make diamond a demanding material for scalable operation. In addition, the $NV^-$ center emits just $\sim5\%$ of its radiation in the ZPL\cite{alkauskas2014first}, limiting significantly the brightness of the source and requiring enhancement of the fraction of photons emitted in the ZPL.

Alternative solutions have been pursued on more fabrication friendly materials that still possess some of diamond's advantages. Silicon Carbide has been proven to be between the most promising platform as an alternative to diamond\cite{weber2010quantum}. SiC is a wide bandgap material $(2.36-3.23eV)$ with relatively high refractive index $(n\approx2.6)$ while at the same time hosts color centers emitting in the near-infrared.
SiC has an already well established fabrication technology, in particular for CMOS high power electronics. The hexagonal polytypes (4H, 6H) exhibit high material purity as they are usually grown using homoepitaxial techniques. In nanophotonics though, thin films are required for high confinement.  For this purpose sophisticated fabrication techniques have to be used for thin film production, including smart cut and electrochemical etching. On the other hand, heteroepitaxial techniques have been developed commercially, offering the access to wafers of cubic (3C) SiC on Si substrates.

Both cubic and hexagonal polytypes host a variety of color centers with properties similar to those of the $NV^-$, with the divacancy ($V_{Si}V_{C}$ one silicon and one neighboring carbon vacancy in SiC lattice) being between the most studied. The presence of divacancy centers has been demonstrated in all three polytypes\cite{falk2013polytype}, showing millisecond coherence times in isolated electron spins\cite{christle2014isolated,christle2017isolated} and coherent control up to room temperature\cite{koehl2011room}. The divacancy spin triplet configuration of the ground and excited states (S=1) and a shelving state that can be addressed for spin initialization makes for a configuration similar to that of the $NV^-$ in diamond. The photonic properties are also similar, with a Debye-Waller factor of $\sim7\%$ and the optical lifetime of about $\sim19$ ns\cite{christle2017isolated}.

To use those color centers for quantum technologies, many schemes rely on the generation of pure, indistinguishable photons, requiring that only photons emitted in the ZPL are collected. The small Debye-Waller factor poses a severe limitation that can be overcome by enhancing the ZPL intensity (hence the portion of photons emitted in it) and lifetime (consequently increasing the count rate, leading to brighter sources). Placing an emitter inside a cavity leads to an enhancement of the spontaneous emission (SE) rate of the emitter expressed by the Purcell factor
\begin{gather}\label{eq:purcell}
F_{P}= \dfrac{\gamma_{en}}{\gamma_{0}}= \dfrac{3}{4\pi^2}\left( \dfrac{\lambda_0}{n} \right)^3 \dfrac{Q}{V}\ \xi
\end{gather}
with $\gamma_{en}$ being the enhanced by the cavity SE rate, $\gamma_0$ the SE rate of the defect in bulk material, $\lambda_0$ the resonant wavelength of the cavity mode, $n$ the refractive index of the material and $Q/V$ the ratio of the cavity's quality factor over the cavity's mode volume. The parameter $\xi$ $\in[0,1]$ describes the overlap of the defect's dipole with the cavity mode.

\begin{figure}[htbp]
\includegraphics[width=0.9\linewidth]{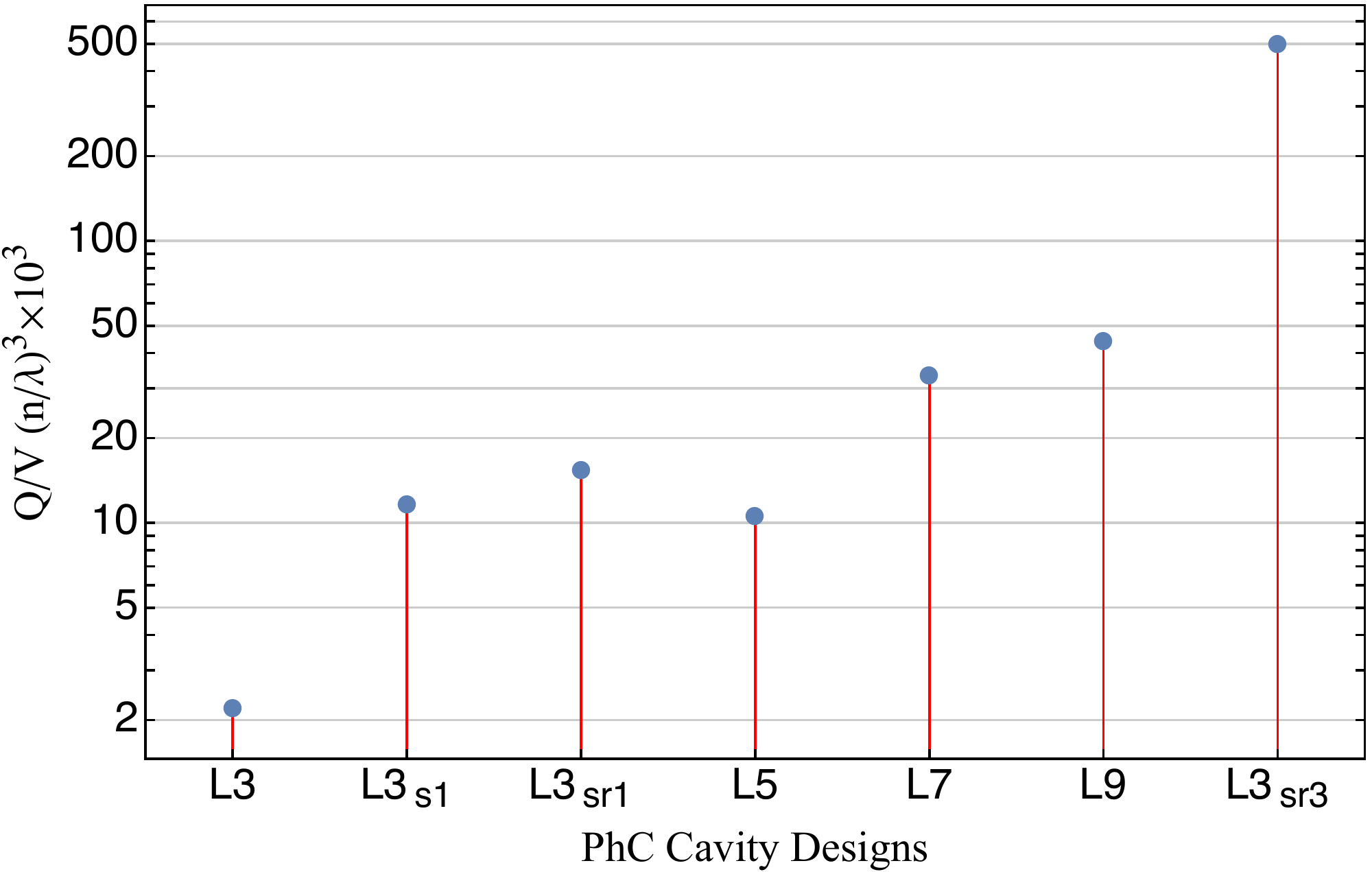}
\caption{The simulated $Q/V$ ratio for various PhCC designs in SiC (n=2.6). All designs are inducing a linear defect (L) to the photonic crystal. The $x$ in L$x$ denotes the number of holes omitted, while $s,r,k$ in L$3_{srk}$ denote that the optimized L3 design includes shifting of holes, reduction of their diameters and the number of holes  undergoing modification next to the cavity, respectively.}
\label{fig:1_Q/V}
\end{figure}

Various attempts of fabricating photonic cavities with high $Q/V$ that would enhance light-matter interaction have been made so far in SiC \cite{song2011demonstration,yamada2012suppression,radulaski2013photonic,bracher2015fabrication,bracher2017selective}, spanning the visible to near-IR, including the telecommunication spectrum. However, limited performances have been demonstrated for cavities operating at the divacancy's ZPL wavelength, showing a maximum Q=1500 in 3C SiC \cite{calusine2014silicon,calusine2016cavity}.
In this Letter we report the simulation and fabrication of optimized L3 photonic crystal cavities (PhCCs) in 3C SiC. Simulated results show a $Q/V \sim 500,000\ (n/\lambda)^3$, while the samples fabricated were measured to have Q factors as high as 7,100.

\section{Simulation}\label{sec:2_sim}
The Purcell factor $F_P$ describes the enhancement of light-matter interaction in a cavity-emitter system and is linked to the brightness and quality of the photons produced by the emitter. As can be seen from equation \ref{eq:purcell}, $F_P$ can be maximized by increasing the $Q/V$ ratio. Longer cavities possess higher $Q$ factors as the portion of the field at the edges, where most of the scattering happens, is decreased. However, longer cavities possess larger modal volume $V$, leading to a small or no gain for the $Q/V$ ratio.
\begin{figure}[htbp]
\includegraphics[width=0.8\linewidth]{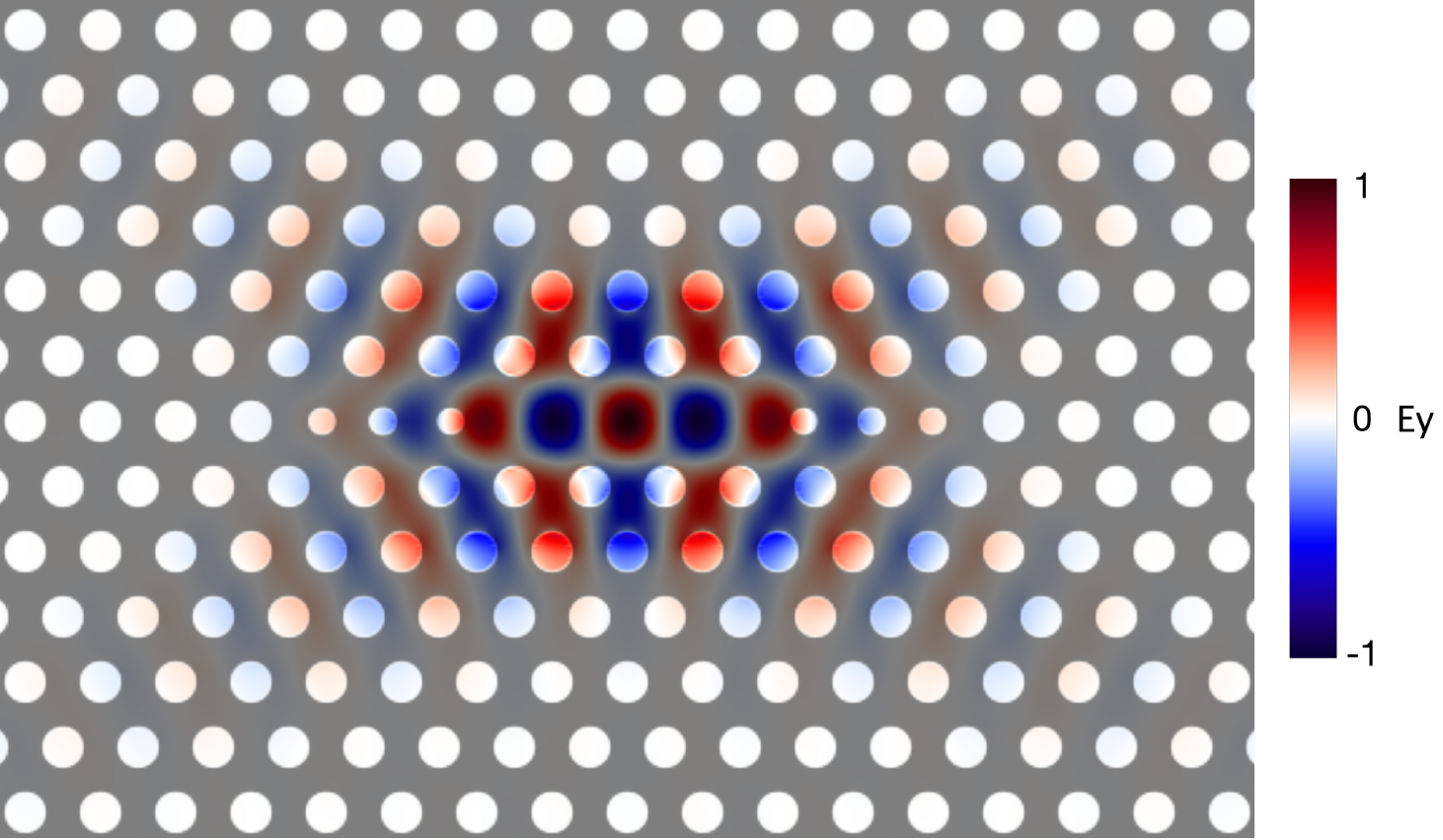}
\caption{The fundamental mode of the simulated Ey component of the electric field in the optimized L$3_{sr3}$ cavity design in SiC. The parameters of the design are; $r/a=0.2553$, $d/a=0.6$, $n=2.6$, $ds_1/a=0.3482$, $ds_2/a=0.2476$, $ds_3/a=0.0573$, $dr_1/a=0.098$, $dr_2/a=0.0882$ and $dr_3/a=0.0927$. The simulated Q factor of this mode is found to be $\sim610,000$ while the modal volume $V\sim1.22 (\lambda/n)^3$.}
\label{fig:2_meep}
\end{figure}
Figure \ref{fig:1_Q/V} shows a comparison of the simulated $Q/V$ ratio for common PhCC designs by 3D finite-difference time-domain (FDTD) in Meep\cite{oskooi2010meep}. Increasing the number of missing holes in the cavity from L3 to L9 produces a sizable increase of the $Q$ factor, so that longer cavities show higher $Q/V$. This trend reaches a plateau as the decrease of the mode at the edge of the cavity does not compensate the increase of modal volume. An alternative approach to maximize $Q/V$ is to optimize the design of the L3 cavity and minimize the scattering losses. As has been described by Noda et. al. work\cite{Akahane2003,Akahane2005}, the optimization of $Q$ in PhCCs is strongly dependent on the manipulation of the cavity mode at the edge of the cavity. Scattering at the edge holes, due to the abrupt transition from the cavity to the photonic crystal, leads to an increase of the radiative modes, increasing the losses and hence decreasing considerably the $Q$ factor. For this reason, a smoother transition from the cavity to the photonic crystal regions should be adopted. 

\begin{figure}[htbp]
\centering
\includegraphics[width=0.7\linewidth]{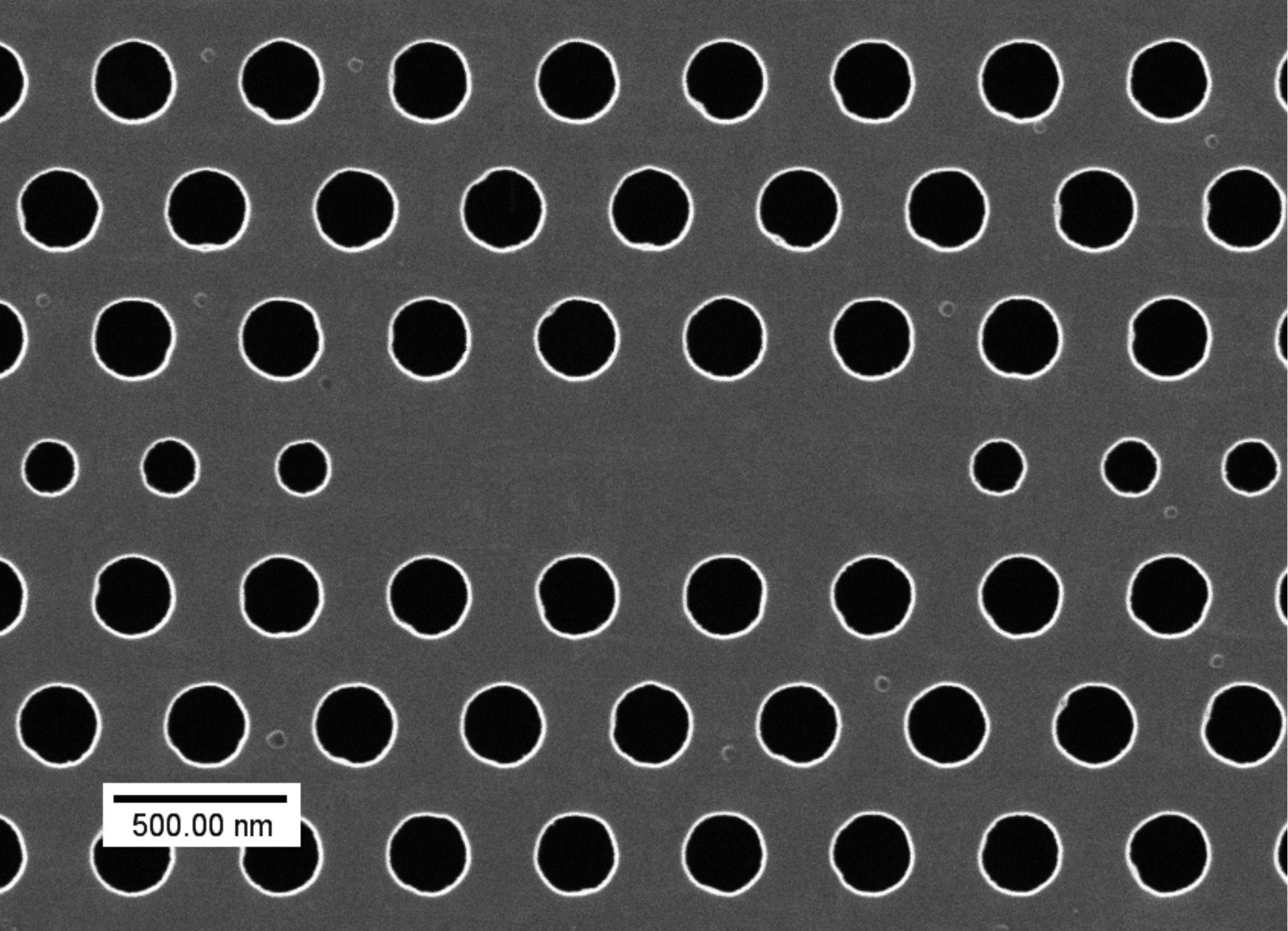}
\caption{Helium Ion Microscope (HIM) image of the fabricated L$3_{sr3}$ PhCC design in 3C SiC.}
\label{fig:3_sample}
\end{figure}

Three different optimized L3 designs are examined. The L$3_{s1}$ design is optimized by shifting the position of one edge hole by 0.21$a$ where $a$ is the lattice constant. Other parameters of the design include $r/a=0.29$ where $r$ the radius of the holes and $d/a=0.6$ where $d$ the slab thickness, while the refractive index used was 2.6. This design shows a simulated $Q$ factor of 9,415 with $V=0.8 (\lambda/n)^3$, compared to 1,429 and $0.64 (\lambda/n)^3$ respectively for the bare L$3$ design. Next L$3_{sr1}$ has one hole which has its position shifted and radius reduced on each side of the cavity. The position shift remains 0.21$a$ whilst the radius is reduced by 0.12$a$. The $Q$ factor of this design is even higher at 14,057 with V$\sim0.9 (\lambda/n)^3$. Finally, the L$3_{sr3}$ is optimized by changing the position and radius of three holes on each side of the cavity. Following similar approach to reference \cite{vico2014gallium}, a radius $r/a=0.2553$ and thickness of the slab $d/a=0.6$ is used. The three holes are shifted by $0.3482a$, $0.2476a$ and $0.0573a$ going closer to further from cavity edge respectively and their radii are reduced by $0.098a$, $0.0882a$ and $0.0927a$. This design gives a quality factor $Q\sim610,000$ and $V\sim1.22 (\lambda/n)^3$. Figure \ref{fig:1_Q/V} shows the design $Q/V$ for the optimized and bare cavities, highlighting how trading off a smaller modal volume can lead to several orders of magnitude higher $Q/V$ ratio. The simulated fundamental mode (Ey component) of the L$3_{sr3}$ cavity can be seen in Figure \ref{fig:2_meep}; it is possible to notice that the optimized holes are placed where the field intensity is close to zero, minimizing scattering whilst producing high confinement.

\section{Results}\label{sec:3_Res}
The optimized cavity was fabricated starting from a wafer of 3C SiC heteroepitaxially grown on Si $\langle001\rangle$ substrate (NOVASiC) that was thinned down to about 240 nm by dry etching. Next a 100 nm layer of aluminum and 5nm of titanium are deposited as hard mask. The titanium cap layer is used to minimize the formation of aluminum oxide on the mask. CSAR62 resist is spun and exposed with a JEOL JBX-9300FS electron beam lithography system. 
A dry etch step of aluminum transfers the pattern from the resist to the hard mask using a chlorine based ICP-RIE etch. The pattern is then transferred to the SiC layer thanks to an additional high-power, low pressure ICP-RIE in SF$_6$ gas. The hard mask and two dry etches are optimized to achieve almost vertical walls in the photonic crystal holes. This is particularly important to obtain high-$Q$ cavities, as the hole verticality was suggested to be the most important limit in previously demonstrated L3 cavities operating at 1100nm\cite{calusine2014silicon}, due to the mixing between TE and TM modes. SEM analysis suggests an etch angle in the holes higher than 86 degrees. In the last fabrication step part of the Si substrate was removed with a TMAH etch, which leaves a SiC suspended membrane. An example of fabricated L$3_{sr3}$ design can be seen in Figure \ref{fig:3_sample}.

\begin{figure}
\includegraphics[width=\linewidth]{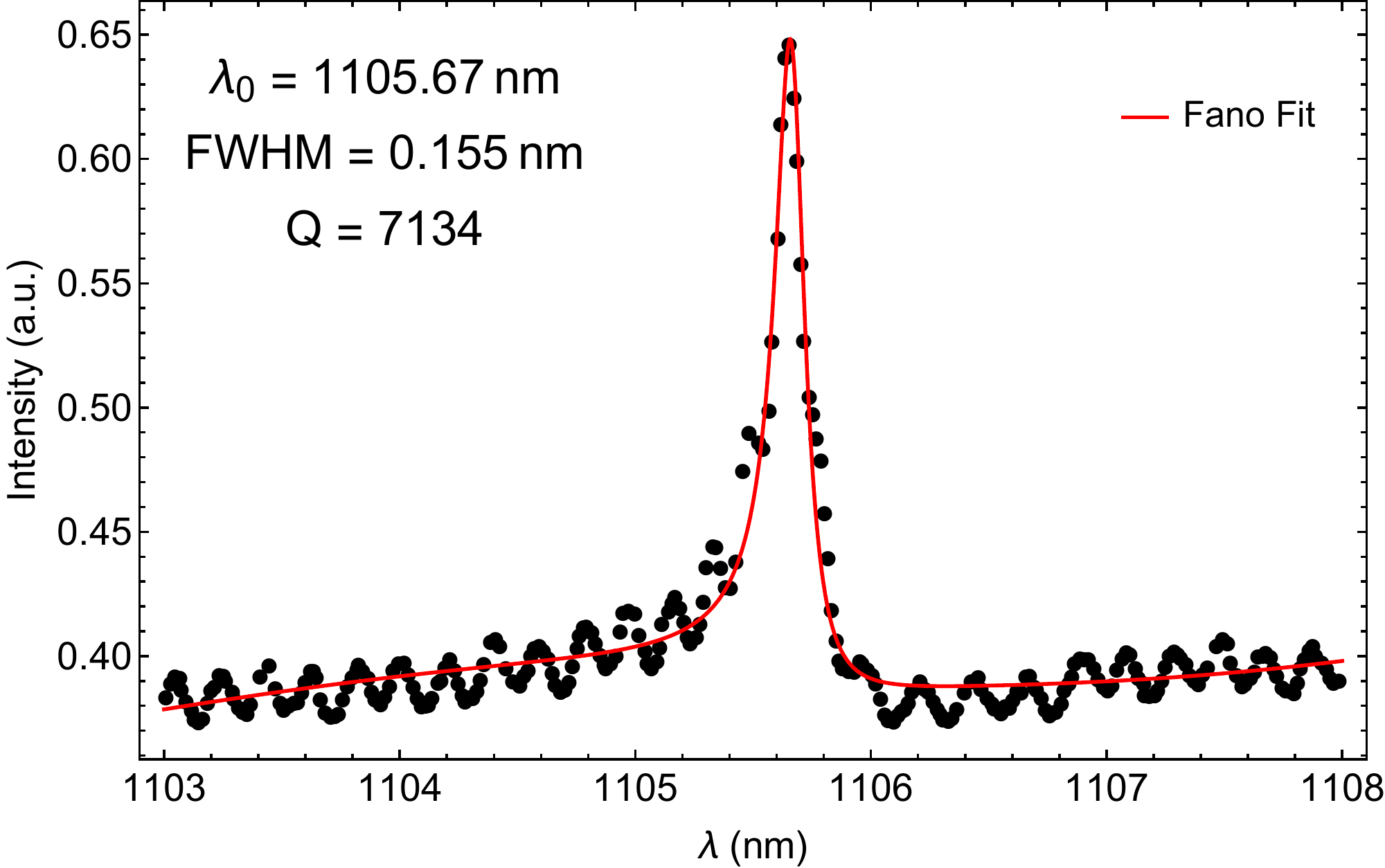}
\caption{Resonant scattering measurement of a cavity similar to this of Figure \ref{fig:3_sample}. A $Q$ factor of 7134 is measured. A Fano type linewidth formula has been used for fitting, a characteristic of resonant scattering technique.}
\label{fig:4_Q}
\end{figure}

The samples are characterized using a home-build confocal microscope setup, equipped a high NA microscope objective (0.85 Olympus). Resonant scattering measurements are performed with an external cavity tunable laser (New Focus TLB-6723, covering the region $1070-1130nm$), which reveals a maximum $Q$ factor of 7,134 (Figure \ref{fig:4_Q}) along with simulated modal volume $V\sim1.22 (\lambda/n)^3$. Although the $Q$ factor measured here is the highest reported for 3C SiC, the measured value is significantly smaller then the simulated one, a problem common in photonic crystal demonstrations. Among the critical parameters responsible for this effect, the sensitivity of the design's $Q$ factor to the optimized holes radii and the thickness of the slab have the biggest impact. 
Another factor to consider is the material losses of 3C SiC\cite{martini2017linear,martini2018four} which are estimated to be high in the defect-rich interface with silicon. In particular, it has been shown that the removal of just 100nm of SiC material close to the silicon interface is sufficient to considerably decrease propagation losses in multimode waveguides\cite{fan2018integrated}. This effect could be even more significant for photonic crystals, as the SiC layer is thinner to support one single mode in the slab.

\section{Discussion}\label{sec:4_Disc}
From the measured $Q$ factor and the simulated modal volume we can estimate that a Purcell factor of 443 could be achieved with the current sample, in the case of color centers perfectly aligned with the cavity ($\xi=1$). Having integrated circuitry and enhanced on chip light-matter interaction is crucial for experiments in quantum optics and quantum information processing. Besides a higher count rate, the Purcell factor also controls two important factors characterizing systems in these fields: the indistinguishability of the photons generated ($I$) and the coupling efficiency of the emitter in the cavity mode ($\beta$). Indistinguishability plays a crucial role in experiments that require two-photon interference and entanglement, while high $\beta$ factor is required for efficient schemes with high emission rates. 

As is described by Iles-Smith et.al.\cite{iles2017phonon}, the presence of the phonon sideband decreases significantly both $I$ and $\beta$, limiting the potentials of solid state emitters, such as quantum dots and defects in solids, for scalable quantum optics. This drawback can be overcome by an emitter-cavity system that shows the adequate Purcell enhancement. For a negligible ratio between the cavity linewidth and the phonon sideband width ($0.155nm/\sim150nm$ in our case), $I$ and $\beta$ can be expressed as,
\begin{gather}\label{eq:Ind}
I=\dfrac{\Gamma_{tot}}{\Gamma_{tot}+2\gamma_{tot}},
\end{gather}
\begin{gather}\label{eq:beta}
\beta=\dfrac{DW~\Gamma_{cav}}{DW~\Gamma_{cav}+\Gamma_{off}}
\end{gather}
\noindent
with, $\Gamma_{tot}$, $\gamma_{tot}$, $DW$, $\Gamma_{cav}$ and $\Gamma_{off}$, being the total emission rate of the emitter, the total dephasing rate, the Debye-Waller factor, the emission rate coupled in the cavity mode and the emission rate not coupled to the cavity mode respectively, while $\Gamma_{tot}=\Gamma_{cav}+\Gamma_{off}$. From the definition of $F_P$ (\ref{eq:purcell}) follows that $\Gamma_{cav}=\gamma_{en}=F_P~\gamma_0$ and in most cases $\Gamma_{off}$ can be approximated by $\gamma_0$. On this basis we can rewrite \ref{eq:Ind} and \ref{eq:beta}
\begin{gather}\label{eq:IndF}
I=\dfrac{F_P+1}{F_P+1+\dfrac{2\gamma_{tot}}{\gamma_0}},
\end{gather}
\begin{gather}\label{eq:betaF}
\beta=\dfrac{F_P}{F_P+1/DW}.
\end{gather}
\indent
Equations \ref{eq:IndF} and \ref{eq:betaF} show the importance of the Purcell factor. Based on the measured $Q\sim 7,134$ and the maximum attainable $F_P\sim443$ the maximum estimated $I$ and $\beta$ values for our system are $0.48$ and $0.97$ respectively by considering the reported dephasing rate\cite{christle2017isolated} of $\sim2$GHz for 3C SiC. Considerably improved performances can be obtained if the dephasing rate can achieve the values reported for 4H polytype of $\sim100$MHz, as they would yield $I\sim0.95$ and $\beta\sim0.97$. These results do not consider the additional benefit of 2D PhCC, as the 2D bandgap ($\sim 135$ nm with the current design) can inhibit unwanted emission in the ZPL through the reduction of the optical density of states\cite{fujita2005simultaneous}.
Higher values of $F_P$ would further enhance $I$ and $\beta$. For an infinite $F_P$ both $I$ and $\beta$ would in principle take near-unity values. In reality, however, a maximum value of $F_P$ exists after which the system cannot be described in the perturbative regime. In the strong coupling condition $I$ starts dropping sharply\cite{iles2017phonon}. 

\begin{figure}[htbp]
\centering
\includegraphics[width=\linewidth]{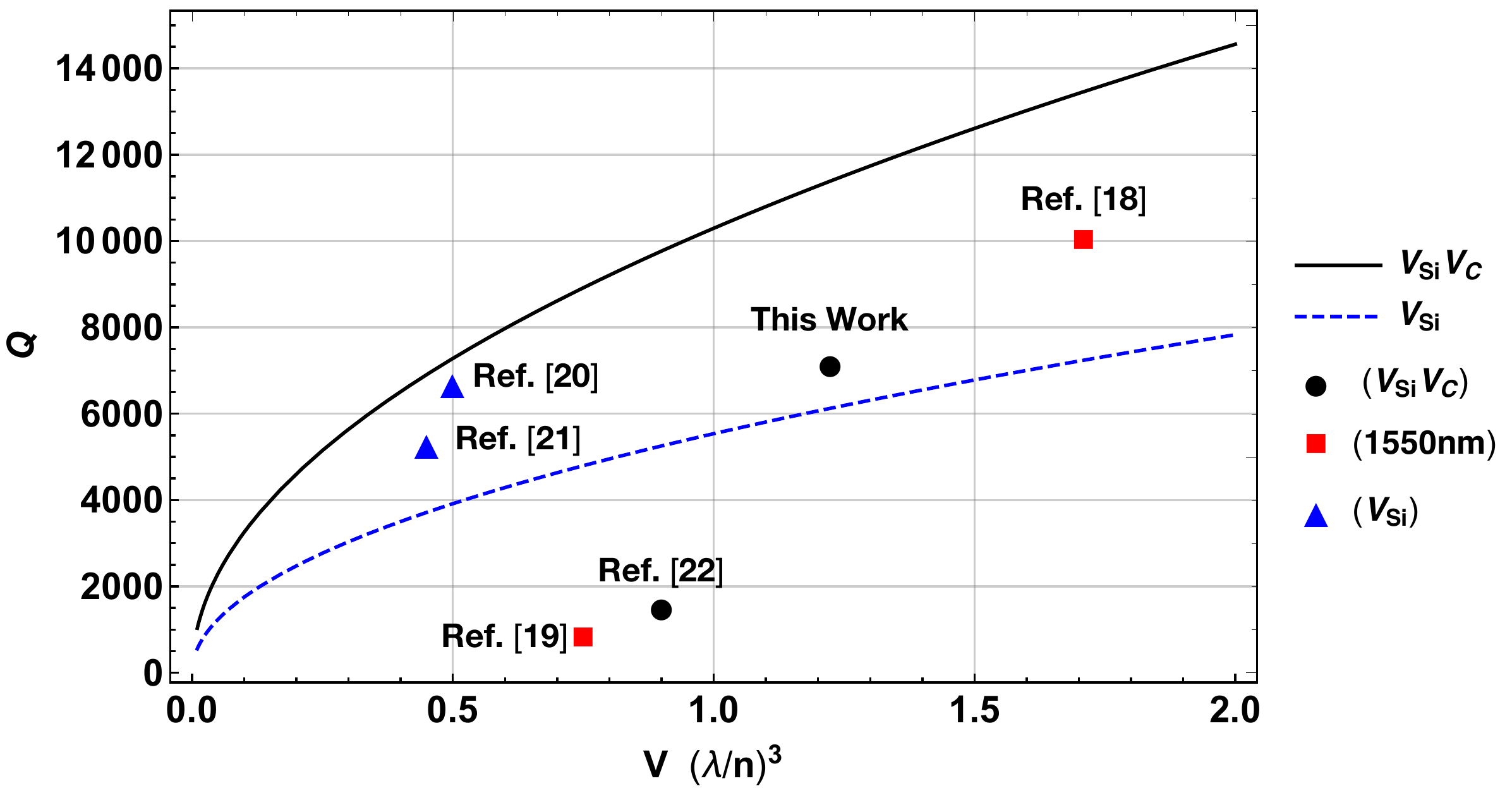}
\caption{Comparison of experimentally demonstrated SiC photonic crystal cavities. The solid black (dashed blue) line represents the strong coupling threshold for the divacancy (silicon vacancy) in SiC. Black dots denote efforts done in divacancy's spectral vicinity, with both of them being at 3C SiC. Red rectangle denote work done in telecom wavelengths with Ref.\cite{yamada2012suppression} being at 6H SiC while Ref.\cite{radulaski2013photonic} at 3C SiC. Blue triangles denote efforts done in the vicinity of the silicon vacancy in 4H SiC.}
\label{fig:5_Str}
\end{figure}

The strong coupling regime threshold is usually defined as $g>|\kappa-\gamma_0|/4$ (see Ref.\cite{andreani1999strong}), where $g$ denotes the emitter-cavity coupling strength, $\kappa=\omega/Q$ the cavity linewidth and $\gamma_0$ the bulk emitter linewidth. Figure \ref{fig:5_Str} shows the strong coupling condition as $Q$ and $V$ are changed for SiC color centers compared with experimental results, and considering that for color centers $\kappa >> \gamma_0$. The figure includes results for both the divacancy as well for the silicon vacancy in SiC. For this defect, single spin manipulation at room temperature has been also demonstrated\cite{Widmann2015coherent} as well as magnetometry and thermometry\cite{kraus2014magnetic}. 
The solid black line (dashed blue line) represent the strong coupling threshold for the divacancy (silicon vacancy). With blue are denoted the efforts done in 4H SiC at the spectral vicinity of the silicon vacancy, with black efforts done in the vicinity of the divacancy and with red the efforts in telecom wavelengths where no defects so far are known in SiC. It's worth noting that the divacancy's strong coupling threshold lies higher than the silicon vacancy one, meaning that larger Purcell factors can be obtained. While Ref.\cite{bracher2015fabrication} lies above the strong coupling line for 4H silicon vacancies, the cavity studied is not coupled to emitters, and Ref.\cite{bracher2017selective} exploited a lower $Q$ factor cavity (2,200) which lies below the strong coupling threshold.

Further optimization in the fabrication of the L$3_{sr3}$ cavity described here to a $Q\sim 11,000$ would put the system at the onset of strong coupling. In this condition the attainable Purcell factor would be 707, corresponding to an increase of $I=0.6$ and $\beta=0.98$ (considering $\gamma_{tot}\sim2$GHz for 3C). This highlights that an improvement of the material is required, even with the optimal cavity parameters, if indistinguishable photons are required.

\section{Conclusion}
In this work we have shown, a new PhCC design which yields $Q/V\sim 500,000 (n/\lambda)^3$ for SiC devices by shifting the position of three holes on the edge of the cavity and reducing their sizes. Samples fabricated in 3C SiC delivered a $Q$ factor of 7,134 which can potentially yield high Purcell factors for enhancing light-matter interaction on chip. Improved values of indistinguishability (currently 0.48 for 3C and 0.95 for 4H) and coupling efficiency (0.97) can be achieved 
for divacancies optimally positioned in the cavity. These values can be further increased with additional improvements in the fabrication and material quality of the SiC layer \cite{fan2018integrated}. Indistinguishability is capped at a maximum of $I=0.6$ by the current value of dephasing in 3C material, as a further increase of Q factor would put the system in the strong coupling regime. Even if this condition limits the performance of the divacancy as single photon source, it would provide access to non-linear effects at the single photon level that can be used for deterministic schemes in quantum computing and simulation. This would make SiC the first scalable platform to exploit color centers in both the weak and strong coupling conditions. The ability to control the regimes of light-matter coupling will pave the way for future quantum technologies in SiC.

\section{Acknowledgement}
We acknowledge support from the Southampton Nanofabrication Centre. This work was supported by the Engineering and Physical Sciences Research Council (EPSRC) EP/P003710/1 and from the European Union Horizon 2020 research and innovation programme under the Marie Sklodowska-Curie grant agreement No 795923.


\end{document}